\documentclass{acm_proc_article-sp}
\def\C {\ensuremath{\mathbb{C}}}

\def\K {\ensuremath{\mathbf{k}}}

\def\Q {\ensuremath{\mathbb{Q}}}
\def\R {\ensuremath{\mathbb{R}}}

\def\KK {\ensuremath{\mathbf{K}}}
\def\u {\ensuremath{\mathbf{u}}}

\def\y {\ensuremath{\mathbf{y}}}
\newtheorem{Theorem}{Theorem}

\newtheorem{Lemma}{Lemma}
\newtheorem{Corollary}{Corollary}

\newdef{Notation}{Notation}
\newdef{Definition}{Definition}
\newdef{Remark}{Remark}
\newdef{Example}{Example}

\newcommand{\init}[1]{\mbox{{\rm init}$(#1)$}}

\newcommand{\mdeg}[1]{\mbox{{\rm mdeg}$(#1)$}}
\newcommand{\mvar}[1]{\mbox{{\rm mvar}$(#1)$}}

\newcommand{\rank}[1]{\mbox{{\rm rank}$(#1)$}}
\newcommand{\res}[1]{\mbox{{\rm res}$(#1)$}}
\newcommand{\ires}[1]{\mbox{{\rm ires}$(#1)$}}
\newcommand{\sat}[1]{\mbox{{\rm sat}$(#1)$}}
\newcommand{\sep}[1]{\mbox{{\rm sep}$(#1)$}}

\newcommand{\Intersect}{{\sf Intersect}}

\newcommand{\MPD}{{\small \sf MPD}}
\newcommand{\SMPD}{{\small \sf SMPD}}
\newcommand{\ctd}{{\scriptsize CTD}}

\newcommand{\Maple}{{\sc Maple}}
\newcommand{\Modpn}{{\sc Modpn}}

\newcommand{\RegularChains}{{\tt RegularChains}}

\newcommand{\cad}{{\scriptsize CAD}}
\newcommand{\CAD}{{\small \sf CAD}}
\newcommand{\Decompose}{{\sf CylindricalDecompose}}
\newcommand{\MakeCylindrical}{{\sf MakeCylindrical}}
\newcommand{\MakeSemiAlgebraic}{{\sf MakeSemiAlgebraic}}
\newcommand{\SeparateZeros}{{\sf SeparateZeros}}
\newcommand{\IsolateZeros}{{\sf IsolateZeros}}
\newcommand{\GenerateStack}{{\sf GenerateStack}}
\newcommand{\Partition}{{\sf InitialPartition}}
\newcommand{\QE}{{\small QE}}

\def\BB {\ensuremath{\mathcal{B}}}
\def\CC {\ensuremath{\mathcal{C}}}
\def\DD {\ensuremath{\mathcal{D}}}
\def\EE {\ensuremath{\mathcal{E}}}
\def\II {\ensuremath{\mathcal{I}}}
\def\JJ {\ensuremath{\mathcal{J}}}
\def\SS {\ensuremath{\mathcal{S}}}
\def\RR {\ensuremath{\mathcal{R}}}
\def\PP {\ensuremath{\mathcal{P}}}
\def\QQ {\ensuremath{\mathcal{Q}}}
\def\TT {\ensuremath{\mathcal{T}}}
\def\cf {\ensuremath{\mathrm{coeff}}}

\def\more-auths{
\end{tabular}
\begin{tabular}{c}}

\title{Computing Cylindrical Algebraic Decomposition via Triangular Decomposition}

\numberofauthors{2}
\author{
\alignauthor Changbo Chen \\
       \affaddr{ORCCA, University of Western Ontario (UWO)} \\
       \affaddr{London, Ontario, Canada} \\
       \email{cchen252@csd.uwo.ca}
\alignauthor Marc Moreno Maza \\
       \affaddr{ORCCA, University of Western Ontario (UWO)} \\
       \affaddr{London, Ontario, Canada} \\
       \email{moreno@csd.uwo.ca}
\vskip1pc
\more-auths
\alignauthor Bican Xia \\
       \affaddr{School of Mathematical Sciences} \\
       \affaddr{Peking University, Beijing, China}
       \email{xbc@math.pku.edu.cn}
\alignauthor Lu Yang \\
       \affaddr{Shanghai Key Laboratory of Trustworthy Computing} \\
       \affaddr{ East China Normal University, Shanghai, China} \\
       \email{ lyang@sei.ecnu.edu.cn}
}

\begin{document}

\maketitle

\begin{abstract}
Cylindrical algebraic decomposition is one of the most
important tools for computing with semi-algebraic sets,
while triangular decomposition is among the most important
approaches for manipulating constructible sets. In this paper,
for an arbitrary finite set $F \subset {\R}[y_1, \ldots, y_n]$
we apply comprehensive triangular decomposition in order to obtain an 
$F$-invariant cylindrical decomposition of
the $n$-dimensional complex space, from which we extract 
an $F$-invariant cylindrical
algebraic decomposition of the $n$-dimensional real spa\-ce.
We report on an implementation of this new approach for constructing 
cylindrical algebraic decompositions.
\end{abstract}

\section{Introduction} %
\label{sect:Introduction}

Cylindrical algebraic decomposition ({\cad}) is a fundamental and powerful
tool in real algebraic geometry.
The original algorithm introduced by Collins in 1973~\cite{col75}
has been followed by many substantial ameliorations, including 
adjacency and clustering techniques~\cite{Arnon84b},
improved projection methods~\cite{scott88,hong90,scott98,brown01}, 
partially built {\cad}s~\cite{ch91,scott93,adam00},
improved stack construction~\cite{Collins02} and 
efficient projection orders~\cite{Dolzmann04}.

The main application of {\cad} is quantifier elimination ({\QE})
for which other approaches are also available.
Some of them have more attractive complexity
results~\cite{bpr06} than {\cad}.
However, as pointed out by Brown and Davenport in~\cite{BrownDavenport2007},
``there is the issue of whether the asymptotic cross-over points
between {\cad} and those other  {\QE} algorithms actually occur in the range
of problems that are even close to accessible with current machines''.
In addition, these authors observe that {\cad} can help solving
certain {\QE} problems~\cite{Brown2001,Hong92}
that other {\QE} algorithms can not.

For a finite set $F_n \subset {\R}[y_1, \ldots, y_n]$ 
the {\cad} algorithm~\cite{col75} decomposes
the real $n$-dimensional space into disjoint cells 
$C_1, \ldots, C_e$ together with  one {\em sample point} $S_i \in C_i$,
for all $1 \leq i \leq e$, such that the sign of each $f \in F_n$ does not change
in $C_i$ and can be determined at $S_i$.
Besides, this decomposition is {\em cylindrical} in the following sense:
For all $1 \leq j < n$ the projections on the first $j$ coordinates $(y_1, \ldots, y_j)$
of any two cells are either disjoint or equal.
We will make use of this notion of ``cylindrical'' decomposition in $\C^n$.

The algorithm of Collins is based on a {\em projection and lifting} procedure
which computes from $F_n$ a finite set 
$F_{n-1} \subset {\R}[y_1, \ldots, y_{n-1}]$ 
such that an $F_n$-invariant {\cad} of ${\R}^n$
can be constructed from an $F_{n-1}$-invariant {\cad} of ${\R}^{n-1}$.
This construction and the base case $n =1$ rely on real root
isolation of univariate polynomials.

In this paper, we propose a different approach 
for computing {\cad}, which proceeds
by successive transformation of an initial decomposition
of the complex $n$-dimensional space.
Our algorithm consists of three main steps:
\begin{description}
\item[{\em \sf Initial Partition:}]  we decompose ${\C}^n$ into
      disjoint constructible sets $C_1, \ldots, C_e$
      such that for all $1 \leq i \leq e$, for each $f \in F_n$ 
       either $f$ is identically zero in $C_i$ or $f$ vanishes 
       at no points of $C_i$.
\item[{\em \sf Make Cylindrical:}] we transform the initial
         partition and obtain another decomposition of ${\C}^n$ into
      disjoint constructible sets such that this second decomposition
      is cylindrical in the above sense.
\item[{\em \sf Make Semi-Algebraic:}] from the previous decomposition
      we produce  an $F_n$-invariant {\cad} of ${\R}^n$.
\end{description}

Our first motivation is to understand the relation
and possible interaction between cylindrical algebraic decompositions
 and triangular decompositions of polynomial systems.
This latter kind of decompositions have been intensively
studied since the work of Wu~\cite{Wu87}.
The papers~\cite{ALM97,BLM06,Hubert03}
and book~\cite{Wang98c} contain 
surveys of the subject.
The primary goal of triangular decompositions is to provide unmixed
decompositions of algebraic varieties.
However, the third and fourth authors have initiated the use
of triangular decompositions in real algebraic geometry~\cite{yhx01}.
Moreover, real root isolation of zero-dimensional polynomial systems
can be achieved via triangular decompositions~\cite{XY02,xz06,Gao07}.

A second motivation of this work is to investigate the possibility of improving
the practical efficiency of {\cad} implementation by means 
of modular methods and fast polynomial arithmetic.
Such techniques have been successfully introduced
into triangular decomposition methods~\cite{DMSWX05a,LMS07,LPM09}.
Each of the three main steps of the algorithm proposed in this paper 
relies on existing sub-algorithms for triangular decompositions
taken from~\cite{MMM99,CGLMP07,xz06} and for which efficient 
implementation in the {\RegularChains} library~\cite{LeMoXi05}
is work in progress based on the highly optimized
low-level routines of the {\Modpn} library~\cite{LMRS08}.

Our third motivation is to extend to real algebraic geometry
the concept of {\em Comprehensive Triangular Decomposition} ({\ctd})
introduced in \cite{CGLMP07}.
The relation between {\cad} and parametric polynomial system
solving is natural as pointed in \cite{Dolzmann98}
and the presentation therein of Weispfenning's approach~\cite{Weispfenning93} 
for {\QE} based on comprehensive Gr\"obner bases.
This suggests that the algorithm proposed in this paper
could support a similar {\QE} method.

This paper is organized as follows.
A summary of the theory of triangular decomposition is given in 
Section~\ref{sect:preliminary}.
Section~\ref{sec:ctd} and Section~\ref{sec:cs}
are dedicated to the first two main steps of our algorithm
whereas Sections~\ref{sec:cad}
presents the last one.
In Section~\ref{sec:experimental} we report on a preliminary
experimentation of our new algorithm.
No modular methods or fast polynomial arithmetic are 
being used yet and our code is just high-level {\Maple}
interpreted code.
However our code can already process well-known examples
from the literature.
We also analyze the performances of the different
main steps and subroutines of our algorithm and implementation.
This suggests that there is a large potential for improvement
by means of modular methods, for instance for the computation
of {\small GCD}s, resultants (and the discriminants)
of polynomials modulo regular chains.

\section{Triangular Decomposition}
\label{sect:preliminary}
Throughout this paper let $\K$ be a field of characteristic zero and
$\KK$ be the algebraic closure of $\K$.
Let $\K[\y]:=\K[y_1,\ldots,y_n]$ be the polynomial ring over the
field $\K$ in variables $y_1 < \cdots < y_n$. Let $p\in {\K}[\y]$ be
a non-constant polynomial. The greatest variable appearing in $p$ is
called the {\em main variable}, denoted by $\mvar{p}$. 
The integer $k$ such that $y_k=\mvar{p}$ is called the {\em level} of $p$.
The {\em}
{\em separant} $\sep{p}$ of $p$ w.r.t $\mvar{p}$, is $\partial
p/\partial\mvar{p}$. The leading coefficient and the
leading monomial of $p$ regarded as a univariate polynomial in
$\mvar{p}$ are called the {\em initial} and
the {\em rank} of $p$; they are denoted by $\init{p}$
and $\rank{p}$ respectively. Let $q$ be another polynomial
of $\K[\y]$, we say $\rank{p}$ is less than $\rank{q}$ if 
$\mvar{p}<\mvar{q}$, or $\mdeg{p}<\mdeg{q}$ when $\mvar{p}=\mvar{q}$.

Let $F\subset \K[\y]$ be a finite polynomial set.
Denote by $\langle F \rangle$ the ideal it generates in
${\K}[\y]$.
Let $h$ be a polynomial in $\K[\y]$,
the {\em{saturated ideal}} $\langle F \rangle:h^{\infty}$
of $\langle F \rangle$ w.r.t $h$, is the set
$\{q\in \K[\y]\mid \exists m\in\mathbb{N}\text{ s.t. }h^mq\in\langle F \rangle\},$
which is an ideal in $\K[\y]$.
The polynomial is {\em regular} modulo $\langle F \rangle$ if it is neither
zero, nor a zerodivisor modulo $\langle F \rangle$.
Denote by $V(F)$ the {\em zero set} (or algebraic variety) of $F$ in ${\KK}^n$.

Let $T \subset {\K}[\y]$ be a {\em triangular set},
that is a set of non-constant polynomials with pairwise
distinct main variables. We denote by $\mvar{T}$ the set of
main variables of polynomials in $T$. A variable in $\y$ is called
{\em algebraic} w.r.t. $T$ if it belongs to \mvar{T}, otherwise
it is called {\em free} w.r.t. $T$. 
For a variable $v \in \y$,
we denote by $T_{<v}$ the subsets of $T$ consisting of the 
polynomials $t$ with main variable less than $v$.
Let $h_T$ be the product
of the initials of polynomials in $T$.
We denote by \sat{T} the saturated ideal of $T$: 
if $T$ is empty then \sat{T} is defined as the trivial
ideal $\langle 0 \rangle$, otherwise it is the ideal
$\langle T \rangle:h_{T}^{\infty}$.
The {\em quasi-component} $W(T)$ of $T$
is defined as $V(T) \setminus V(h_T)$.
Let $h\in\K[\y]$ be a polynomial. Define 
$Z(T,h):= \ W(T) \setminus V(h).$

Let $h \in  \K[\y]$ be a polynomial.
The {\em iterated resultant} of $h$ w.r.t. $T$, denoted by
\ires{h,T}, is defined as follows:
$(1)$ if $h \in {\K}$ or all variables in $h$ are free w.r.t. $T$,
      then $\ires{h,T} = h$; 
$(2)$ otherwise, if $v$ is the largest variable of $h$
      which is algebraic w.r.t. $T$, then
      $\ires{h,T} = \ires{r,T_{<v}}$
      where $r$ is the resultant of $h$ and
      the polynomial in $T$ whose main variable is $v$.
Iterated resultants have the following important property:
the polynomial $h$ is regular modulo \sat{T} if and only if
we have $\ires{h, T} \neq 0$.

We say that the triangular set $T$ is a {\em regular chain}
if either $T=\varnothing$ or $\ires{h_T, T} \neq 0$.
The pair $[T,h]$ is called a {\em regular system} if $T$ is a regular chain,
and $\ires{h, T} \neq 0$.
Denote by $\sep{T}$ the product of all $\sep{p}$, where $p\in T$.
Then $T$ is said to be {\em squarefree} if $\ires{\sep{T}, T} \neq 0$.
A regular system $rs=[T,h]$ is said to be {\em squarefree} 
if $T$ is squarefree.

For a regular system $rs=[T,h]$, the rank of $rs$, denoted by $\rank{rs}$,
is defined as the set of $\rank{p}$ for all $p\in T$.
Given another regular system $rs'=[T',h']$
with $\rank{rs}\neq\rank{rs'}$, we say $\rank{rs}$ is less than $\rank{rs'}$ 
whenever the minimal element of the symmetric difference
$(\rank{rs}\setminus\rank{rs'}) \, \cup \, (\rank{rs'}\setminus\rank{rs})$ 
belongs to $\rank{rs}$. 

A {\em constructible set} of $\KK^n$ is
any finite union
$$(A_1 \setminus B_1) \ {\cup} \ \cdots \ {\cup} \ (A_e \setminus B_e)$$
where $A_1, \ldots, A_e, B_1, \ldots, B_e$ are algebraic varieties in $\KK^n$.
For any constructible set $cs$ of $\KK^n$ 
there exist finitely many regular systems 
${rs}_1, \ldots, {rs}_m$ of ${\K}[\y]$ such that
$cs = Z({rs}_1) \ \cup \ \cdots  \ \cup \ Z({rs}_m).$
\begin{Example}
Consider the polynomials in $\K[y_1<y_2<y_3]$
$$
p_1 = y_2^2+y_1-1~\mbox{and}~p_2=y_1y_3^2-1.
$$
We illustrate the previous main notions as follows.
\begin{center}
\fbox{
\mbox{
\begin{tabular}{ccccc}
    & \rm{mvar} & \rm{sep}   & \rm{init} & \rm{rank}\\
$p_1$ & $y_2$   & $2y_2$     & $1$       & $y_2^2$\\
$p_2$ & $y_3$   & $2y_1y_3$  & $y_1$     & $y_1y_3^2$\\
\end{tabular}
}}
\end{center}
The initial $y_1$ of $p_2$ is regular modulo $\langle p_1 \rangle$. 
The set $T=\{p_1,p_2\}$ is a triangular set.
The iterated resultant of $y_1$ and $T$ is $y_1$, 
so $T$ is a regular chain. 
The pair $[T,y_2]$ is a regular system, since $\ires{y_2, T}=y_1-1$. 
The quasi-component of $T$ is the set of points in $\KK^3$ such that
$p_1=0$, $p_2=0$ and $y_1\neq 0$,
which is a constructible set.
\end{Example}
We review three important operations ${\Intersect}$, 
{\tt Make\-Pair\-wise\-Dis\-joint} ({\MPD}) and 
{\tt Sym\-me\-tri\-cal\-ly\-Make\-Pair\-wise\-Dis\-joint} ({\SMPD})
proposed in~\cite{CGLMP07}.
Let $rs_*=[T_*,h_*]$ be a squarefree regular system of $\K[\y]$ and let
$p$ be a polynomial of $\K[\y]$ such that $p$ is regular w.r.t $\sat{T_*}$. 
The operation $\Intersect(p, rs_*)$ computes a family
of squarefree regular systems ${\RR}$ of $\K[\y]$ such that
$$V(p)\cap Z(rs_*)=\cup_{rs\in{\RR}} Z(rs),$$
and the rank of each $rs\in{\RR}$ is less than that of $rs_*$.

For regular systems $[T_1, h_1], \ldots, [T_e, h_e]$ in ${\K}[\y]$,
the function {\MPD}
returns regular systems $[S_1, g_1], \ldots, [S_f, g_f]$ in ${\K}[\y]$
s.t.
$$Z(T_1, h_1) \ \cup \ \cdots \ \cup \ Z(T_e, h_e) \  = \
               Z(S_1, g_1) \ \cup \ \cdots \ \cup \ Z(S_f, g_f),$$
and for all $1 \leq i < j \leq f$ we have
              $Z(S_i, g_i) \ \cap \ Z(S_j, g_j) = \varnothing.$

Given a family $\CC=\{C_1, \ldots, C_r\}$ 
of constructible sets of  $\KK^n$,
the function {\SMPD}
returns a family $\DD=\{D_1,\ldots, D_s\}$
of constructible sets of  $\KK^n$ 
such that $D_i\cap D_j=\varnothing$ for all $1\leq i <  j\leq n$, 
each $D_j$ is a subset of some $C_i$, and each $C_i$ can be written as a finite 
union of some of the $D_j$'s.
Such a $\DD$ is called an {\em intersection-free basis} 
of  $\CC$.

\section{Zero Separation}
\label{sec:ctd}
In this section, we assume $n \geq 2$ 
and we regard the ordered variables $y_1<\cdots<y_{n-1}$
as parameters, denoted by $\u$. 
Let ${\pi}_\u$ be the projection function which
sends a point $(\bar{\u},\bar{y_n})$ of $\KK^{n}$ to the point 
$\bar{\u}$ of the parameter space $\KK^{n-1}$.
Let $\bar{\u} \in \KK^{n-1}$.
We write ${\pi}_\u^{-1}(\bar{\u})$ for the set 
of all points $(\bar{\u},\bar{y_n})$ 
in $\KK^{n}$ such that ${\pi}_\u(\bar{\u},\bar{y_n})=\bar{\u}$.

Let $p \in \K[\u,y_n]$ be a polynomial of level $n$, 
that is, with main variable $y_n$.
In broad terms, 
the goal of this section is to decompose the parameter space
$\KK^{n-1}$ into finitely many cells such that above each cell
the ``root structure'' of $p$ (number of roots, their multiplicity, \ldots)
does not change.
In fact, we make this problem more general by allowing 
algebraic constraints on the parameter $\u$.
After some notations,
we define in Definition~\ref{Defi:WellSeparated}
the object to be computed by the algorithm devised in this section.
It can be seen as a specialization of the 
comprehensive triangular decomposition ({\ctd})
to the case where the input system is a regular system and
all variables but one are regarded as parameters.
This algorithm is stated in Section~\ref{sec:TheAlgorithmSeparateZeros}
after two lemmas.

\smallskip\noindent{\small \bf Notations.}
Let $rs=[T,h]$ be a regular system of $\K[\u,y_n]$. 
If $y_n$ does not appear in $rs$, we denote by $Z_\u(rs)$
the zero set of $rs$ in $\KK^{n-1}$.
If $y_n$ does not appear in $T$, we write $W_\u(T)$ for
the quasi-component of $T$ in $\KK^{n-1}$.
If $\mvar{h}=y_n$ holds, we denote by 
$\cf(h)$ be the set of coefficients of $h$ when $h$ is 
regarded as a polynomial in $y_n$ with coefficients in $\K[\u]$
and by $V_\u(\cf(h))$ the variety of $\cf(h)$ in $\KK^{n-1}$.
Finally, if $y_n$ is algebraic in $T$, 
letting $t_n$ be the polynomial in $T$ with main variable $y_n$,
we write $T_\u=T\setminus\{t_n\}$ and $rs_\u=[T_\u, r]$,
where $r=\res{h\cdot \sep{t_n}, t_n}$ is the resultant
of $h\cdot \sep{t_n}$ and $t_n$ w.r.t $y_n$. 

\begin{Definition}
\label{Defi:WellSeparated}
Let $C$ be a constructible set of $\KK^{n-1}$.
A finite set of level $n$ 
polynomials ${\PP}\subset\K[\u,y_n]$
{\em separates above} $C$ if for each $\alpha\in C$:
$(1)$ the initial of any $p\in{\PP}$ does not vanish at $\alpha$;
$(2)$ the polynomials $p(\alpha,y_n) \in \KK[y_n]$, $p\in{\PP}$, 
              are squarefree and coprime.

Let  ${\CC}$ be a finite collection of pairwise disjoint 
constructible sets of $\KK^{n-1}$, and, for each $C\in{\CC}$,
let ${\PP}_C\subset\K[\u,y_n]$ 
be a finite set of level $n$ polynomials.
Let $rs_*=[T_*,h_*]$ be a regular system of $\K[\u,y_n]$,
where $n \geq 2$ and $y_n$ is algebraic w.r.t $T$.
We say that the family $\{(C, {\PP}_C)\mid C\in{\CC}\}$
{\em separates} $Z(rs_*)$ if the following conditions hold:
\begin{enumerate}
\item[$(1)$] ${\CC}$ is a partition of ${\pi}_\u(Z(rs_*))$,
\item[$(2)$] for each $C\in{\CC}$, ${\PP}_C$ separates above $C$,
\item[$(3)$]
$
Z(rs_*)=\bigcup_{C\in{\CC}}\bigcup_{p\in{\PP}_C} V(p)\cap\pi_\u^{-1}(C).
$
\end{enumerate}
More generally, let $cs$ be a constructible set of ${\KK}^n$
such that there exist regular systems $rs_1, \ldots, rs_r$
of $\K[\u,y_n]$ whose zero sets form a partition
of $cs$ and such that $y_n$ is algebraic w.r.t. the regular chain of $rs_i$,
for all $1 \leq i \leq r$.
Then, we say that the family $\{(C, {\PP}_C)\mid C\in{\CC}\}$
{\em separates} $cs$ if ${\CC}$ is a partition
of ${\pi}_\u(cs)$ and if for all $1 \leq i \leq r$ there exists
a non-empty subset ${\CC}_i$ of ${\CC}$
and for each $C \in {\CC}_i$
a non-empty subset ${\PP}_{C,i} \subseteq {\PP}_C$
such that $\{(C, {\PP}_{C,i})\mid C\in{\CC}_i\}$
separates  $Z(rs_i)$. 
In this case, we have:
$
cs=\bigcup_{C\in{\CC}}\bigcup_{p\in{\PP}_C} V(p)\cap\pi_\u^{-1}(C).
$
\end{Definition}
\begin{Example}
Consider the polynomials in $\K[x>b>a]$
$$
 p_1 = ax^2-b~\mbox{and}~p_2=ax^2+2x+b,
$$
and the constructible set $C=\{(a, b)\in\KK^2\mid ab(ab-1)\neq0\}$.
For any point $(a, b)$ of $C$, the two polynomials $p_1(a, b)$ and $p_2(a, b)$
of $\KK[x]$ are squarefree and coprime. 
So the polynomial set $\{p_1,p_2\}$ separates above $C$.

Consider the regular system $rs_*=[\{p_1\},1]$ and the constructible
sets
$$
\begin{array}{l}
C_1=\{(a, b)\in\KK^2\mid ab\neq0\}\\
C_2=\{(a, b)\in\KK^2\mid a\neq0 ~\&~ b=0\} 
\end{array}
$$
Note that the zero set of $rs_*$ 
is $\{p_1=0~\&~ a\neq0\}$. 
So the family $\{~(C_1,\{p_1\}), ~(C_2,\{ax\})~\}$ separates $Z(rs_*)$.

Given two regular systems
$$
rs_1=[\{p_1\}, b]~\mbox{and}~ rs_2=[\{p_2, b\},1].
$$
Consider the constructible set
$$
\begin{array}{rcl}
cs &=& Z(rs_1)\cup Z(rs_2)\\
   &=& \left(V(p_1)\setminus V(ab)\right) \cup \left(V(p_2,b)\setminus V(a)\right).
\end{array}
$$
The family $\{~(C_1,\{p_1\}), ~(C_2,\{p_2\})~\}$ separates $cs$.
\end{Example}

\begin{Lemma}
\label{Lemma:res2}
Let $p \in \K[\u,y_n]$ be a level $n$ polynomial.
Let $r=\res{\sep{p}, p}$ be the resultant of $\sep{p}$ and $p$ w.r.t $y_n$.
Then, the polynomial $p(\bar{\u})$ of $\KK[y_n]$ is squarefree
and $\init{p}$ 
does not vanish at $\bar{\u}\in\KK^{n-1}$, if and only if,
$r(\bar{\u})\neq0$ holds.
\end{Lemma}
Observe that $\init{p}$ is a factor of $r$. So the conclusion
follows directly from the specialization property of subresultants.

\begin{Lemma}
\label{Lemma:res3}
We have the following properties:
\begin{enumerate}
\item[$(1)$] If $y_n$ does not appear in $rs$, then
             $\pi_\u(Z(rs))=Z_\u(rs)$.
\item[$(2)$] If $y_n$ does not appear in $T$ and if $\mvar{h}=y_n$ holds,
              then we have
                $\pi_\u(Z(rs))=W_\u(T)\setminus V_\u(\cf(h))$.
\item[$(3)$] If $y_n$ is algebraic w.r.t $T$ and if the regular
               system $rs$ is squarefree, then $rs_\u$ 
               is a squarefree regular system of $\K[\u]$;
               moreover there exists a family
               ${\RR}'$ of squarefree regular systems of $\K[\u,y_n]$ such that:
\begin{enumerate}
\item[$(a)$] the rank of each $rs'\in{\RR}'$ is less 
             than that of $rs$,
\item[$(b)$] for each $[T',h']\in{\RR}'$, $y_n$ is algebraic w.r.t $T'$,
\item[$(b)$] the zero sets $Z(rs')$, $rs'\in{\RR}'$ and 
             the zero set $V(t_n) \cap Z(rs_\u)$
             are pairwise disjoint, and we have
\item[$(d)$] 
$Z(rs) = V(t_n) \cap Z(rs_\u) \ \cup \
       \bigcup_{rs'\in{\RR}'} Z(rs').$
\end{enumerate}
\end{enumerate}
\end{Lemma}
\begin{proof}
Property $(1)$ is clear and proving $(2)$ is routine.
We prove $(3)$.
Since $rs$ is squarefree, using the above notations, we have
$$
\ires{r, T}=\ires{r,T_\u}=\ires{h\cdot \sep{t_n}, T}\neq0.
$$
This implies that $r$ is regular w.r.t $\sat{T}$ and that
$rs_\u=[T_\u, r]$ is a squarefree regular system of $\K[\u]$.
Observe now  that the zero set of $rs$ decomposes
in two disjoint parts:
$$
Z(rs)=\left(Z(rs)\setminus V(r)\right)\cup \left(Z(rs)\cap V(r)\right). 
$$
For the first part, we have
$$
Z(rs)\setminus V(r)= V(t_n) \cap Z(rs_\u).
$$
For the second part, since $r$ is regular w.r.t $\sat{T}$,
by calling operation $\Intersect$, we obtain a family
${\RR}$ of squarefree regular systems of $\K[\u,y_n]$
such that
$$
Z(rs)\cap V(r)=\bigcup_{rs'\in{\RR}} Z(rs'),
$$
where the rank of each $rs'\in{\RR}$ is less 
than that of $rs$.
Finally, applying the operation {\MPD} to ${\RR}$
we obtain a family ${\RR}'$ satisfying the
properties $(a)$, $(b)$, $(c)$ and $(d)$.
\end{proof}

\subsection{The Algorithm {\SeparateZeros}} %
\label{sec:TheAlgorithmSeparateZeros}

We present now an algorithm 
``solving'' a regular system
in the sense of Definition~\ref{Defi:WellSeparated}.
Precise specializations and algorithm steps follow.

\noindent\underline{\small \bf Calling sequence.} 
${\SeparateZeros}(rs_*, \u, n)$

\noindent{\small \bf Input.}
A (squarefree) regular system $rs_*=[T_*,h_*]$ of $\K[\u, y_n]$,
where $n \geq 2$ and $y_n$ is algebraic w.r.t $T_*$.

\noindent{\small \bf Output.} A finite
family $\{(C, {\PP}_C)\mid C\in{\CC}\}$, 
where ${\CC}$ is a finite collection of
constructible sets of $\KK^{n-1}$, and for each $C\in{\CC}$,
${\PP}_C\subset\K[y_1,\ldots,y_n]$ 
is a finite set of level $n$ polynomials, 
such that $\{(C, {\PP}_C)\mid C\in{\CC}\}$ 
separates the zero set of $rs_*$.
(See Definition~\ref{Defi:WellSeparated}.)

\noindent{\small \bf Step $(1)$.}
Initialize ${\RR}=\{rs_*\}$ and ${\PP}=\varnothing$.

\noindent{\small \bf Step $(2)$.}
If ${\RR}=\varnothing$, go to {\small \bf Step $(3)$.}
Otherwise
arbitrarily choose one regular system $rs=[T,h]$ from ${\RR}$ 
and let ${\RR}={\RR}\setminus\{rs\}$.
Using the above notations, let ${\RR}'$ be as in Property $(3)$
of Lemma~\ref{Lemma:res3}.
Set ${\PP}={\PP}\cup\{(rs_\u, t_n)\}$, 
set ${\RR}={\RR}\cup{\RR}'$ and repeat {\small \bf Step $(2)$.}

\noindent{\small \bf Comment.}
Observe that Step $(2)$ will finally terminate since each newly 
added regular system into ${\RR}$ has a rank less than that of
the one removed from ${\RR}$.
When Step $(2)$ terminates, we obtain a family ${\PP}$ of pairs such that
$$
Z(rs_*)=\bigcup_{(rs_\u, t_n)\in{\PP}} V(t_n)\cap{\pi}_u^{-1}(Z_\u(rs_\u)),
$$
and the union is disjoint.
Next, observe that for 
 each pair $(rs_\u, t_n)\in{\PP}$, the polynomial $\init{t_n}$ 
does not vanish at  any point of $Z_\u(rs_\u)$, by virtue
of Lemma~\ref{Lemma:res2}.
Therefore, the union of all $Z_\u(rs_\u)$ is equal to ${\pi}_\u(Z(rs_*))$.

\noindent{\small \bf Step $(3)$.}
By means of the operation {\small \sf SMPD} we compute
an intersection-free basis of all $Z_\u(rs_\u)$.
Hence we obtain a partition ${\CC}$ of ${\pi}_\u(Z(rs_*))$.
Then, for each $C \in {\CC}$ we define
${\PP}_C$ as the set of the polynomials $t_n$
such that there exists a regular system $rs_\u$
satisfying $(rs_\u,t_n) \in {\PP}$ and $C \subseteq Z_\u(rs_\u)$.
Clearly $\{(C,{\PP}_C)\mid C\in{\CC}\}$ is a valid output.

Finally, we generalize this algorithm in order to apply it 
to a constructible set represented by regular systems.

\noindent\underline{\small \bf Calling sequence.} 
${\SeparateZeros}(\{rs_1, \ldots, rs_r\} , \u, n)$

\noindent{\small \bf Input.}
Regular systems $rs_1, \ldots, rs_r$
of $\K[\u, y_n]$, $n\geq2$, whose zero sets are pairwise disjoint
and such that $y_n$ is algebraic w.r.t. the regular chain of $rs_i$,
for all $1 \leq i \leq r$; let $cs$ be the constructible set
represented by $rs_1, \ldots, rs_r$.

\noindent{\small \bf Output.} A finite
family $\{(C, {\PP}_C)\mid C\in{\CC}\}$,
where ${\CC}$ is a finite collection of 
constructible sets of $\KK^{n-1}$, and for each $C\in{\CC}$,
${\PP}_C\subset\K[y_1,\ldots,y_n]$ 
is a finite set of level $n$ polynomials, 
such that $\{(C, {\PP}_C)\mid C\in{\CC}\}$ 
separates $cs$.
(See Definition~\ref{Defi:WellSeparated}.)

\noindent{\small \bf Step $(1)$.}
For each $1 \leq i \leq r$, call
${\SeparateZeros}(rs_i, \u, n)$
obtaining 
$\{(C, {\PP}_C)\mid C\in{\CC}_i\}$
where ${\CC}_i$ is a partition of ${\pi}_\u(Z(rs_i))$.

\noindent{\small \bf Step $(2)$.}
By means of the operation {\SMPD}, 
compute an intersection-free basis ${\DD}$
of the union of the  ${\CC}_i$, for $1 \leq i \leq r$.

\noindent{\small \bf Step $(3)$.}
For each $D \in {\DD}$, let ${\PP}_D$ be the union
of the ${\PP}_C$ such that $D \subseteq C$ holds.
Return $\{(D, {\PP}_D) \mid D\in{\DD}\}$.

\section{Cylindrical Decomposition}
\label{sec:cs}
In this section,
we propose the notion of
an {\em $F$-invariant cylindrical decomposition} of $\KK^n$,
generalizing ideas that are well-known in the case of real fields.
The main algorithm and its subroutines
for computing such a decomposition are stated in three subsections.

\begin{Definition}
\label{Defi:CylindricalDecomposition} %
We state
the definition by induction on $n$.
For $n=1$, a cylindrical decomposition of $\KK$ is a finite collection of sets
$\{D_1,\ldots,D_{r+1}\}$, where either $r=0$ and $D_1=\KK$, or $r>0$ and
there exists $r$ 
nonconstant coprime squarefree polynomials $p_1,\ldots,p_{r}$ of $\K[y_1]$ such
that
$$
D_i=\{y_1\in\KK\mid p_i(y_1)=0\}, 1\leq i\leq r,
$$
and $D_{r+1}=\{y_1\in\KK\mid p_1(y_1)\cdots p_r(y_1) \neq 0\}.$
Note that all $D_i$, $1\leq i\leq {r+1}$ form a partition
of $\KK$. 
Now let $n>1$, and let ${\DD}'=\{D_1,\ldots,D_s\}$ be any cylindrical decomposition of $\KK^{n-1}$.
For each $D_i$, 
let $\{p_{i,1},\ldots,p_{i,r_i}\}$, $r_i\geq0$, be a set of polynomials 
which separates above $D_i$. (See Definition~\ref{Defi:WellSeparated}.)
If $r_i=0$, set $D_{i,1}=D_i\times\KK$. 
If $r_i>0$, set
$$
D_{i,j}=\{(\alpha,y_n)\in\KK^n\mid \alpha\in D_i~\&~p_{i,j}(\alpha,y_n)=0\},
$$
for $1\leq j\leq r_i$ and set
$$
D_{i,r_{i}+1}=\{(\alpha,y_n)\in\KK^n\mid \alpha\in D_i~\&~\left(\prod_{j=1}^{r_i}p_{i,j}(\alpha,y_n)\right)\neq0\}.
$$
The collection ${\DD}=\{D_{i,j}\mid 1\leq i\leq s, 1\leq j\leq r_{i}+1\}$ is called a
cylindrical decomposition of $\KK^n$. Moreover, we say that ${\DD}$ induces ${\DD}'$.

Let $F=\{f_1,\ldots,f_s\}$ be a finite set of polynomials of $\K[y_1<\cdots<y_n]$. 
A cylindrical decomposition ${\DD}$ of $\KK^n$ is called {\em $F$-invariant}
if ${\DD}$ is an intersection-free basis of the $s+1$ constructible
sets $V(f_i),$ $1\leq i\leq s$ and 
$\{y\in\KK^n \mid f_1(y)\cdots f_s(y) \neq 0 \}.$
\end{Definition}

\begin{Lemma}
\label{Lemma:res4}
Let $rs_1,\ldots, rs_{r+1}$, with $r\geq1$,
be regular systems of ${\K}[y_1]$ 
such that their zero sets  form a partition of ${\KK}^1$.
Then, up to renumbering, there
exist polynomials $p_1, \ldots, p_r$, $
h_1, \ldots, h_r, h_{r+1} \in {\K}[y_1]$
such that $rs_i = [ \{ p_i \}, h_i ]$ for $1 \leq i \leq r$
and $rs_{r+1} = [ \varnothing, h_{r+1}]$.
Moreover, setting
    $D_i = V(p_i)$ for $1 \leq i \leq r$
and $D_{r+1} = \{ y_1 \in \KK \mid  p_1(y_1) \cdots p_r(y_1) \neq 0 \}$,
the sets $D_1, \ldots, D_{r+1}$ form a cylindrical decomposition of $\KK$.
\end{Lemma}
\begin{proof}
Observe that for $1 \leq i \leq r$
we have $Z(rs_i) = V(p_i)$, as $h_i$ and $p_i$ have no common roots.
Since the zero sets $Z(rs_1),\ldots, Z(rs_{r+1})$ 
form a partition of ${\KK}^1$,
we must have $V(h_{r+1}) = V(p_1 \cdots p_r)$.
The conclusion follows.
\end{proof}

\subsection{The Algorithm {\MakeCylindrical} } %
\label{Sect:TheAlgorithmMakeCylindrical}

\noindent\underline{\small \bf Calling sequence.} 
${\MakeCylindrical}({\RR},n)$

\noindent{\small \bf Input.} ${\RR}$, a finite family
        of regular systems such that the zero sets 
        $Z(rs)$, for all $rs\in{\RR}$, form a partition of $\KK^n$.

\noindent{\small \bf Output.} ${\DD}$, a cylindrical 
  decomposition of $\KK^n$ 
  such that the zero set of each regular system in ${\RR}$ 
  is a union of some cells in ${\DD}$.

\noindent{\small \bf Step $(1)$: {Base case}.}
If $n>1$, go to $(2)$. If ${\RR}$ has only one element, 
return ${\DD}=\KK$ otherwise use the construction
of Lemma~\ref{Lemma:res4} to return a
cylindrical decomposition ${\DD}$.

\noindent{\small \bf Step $(2)$: {Initialization}.}
Set to ${\RR}_1, {\RR}_2, {\RR}_3$
the subset of ${\RR}$ consisting of regular systems $rs=[T,h]$ such
that, $y_n$ is algebraic w.r.t $T$, 
     $y_n$ appears in $h$ but not in $T$,
     $y_n$ does not appear in $T$ nor in $h$, respectively.

\noindent{\small \bf Step $(3)$: {Processing ${\RR}_1$}.}
Call ${\SeparateZeros}({\RR}_1,\u, n)$
(see  Section~\ref{sec:ctd})
obtaining 
$\{(C, {\PP}_C)\mid C \in {\CC}_1 \}$ 
where ${\CC}_1$ is a partition of ${\pi}_\u(cs_1)$,
where $cs_1$ is the constructible set represented by ${\RR}_1$.
By adding a ``$1$'' in each pair, we obtain a collection of triples
${\TT}_1=\{(C, {\PP}_{C}, 1)\mid C\in{\CC}_1\}$.

\noindent{\small \bf Step $(4)$: {Processing ${\RR}_2$}.}
For each $rs\in {\RR}_2$, 
compute the projection $\pi_\u(Z(rs))$ by Property $(2)$ of Lemma~\ref{Lemma:res3}. 
Set ${\CC}_2 = \{   \pi_\u(Z(rs)) \ \mid \  rs\in {\RR}_2\}$
and
${\TT}_2=\{(C, \varnothing, 2)\mid C\in{\CC}_2\}$.

\noindent{\small \bf Step $(5)$: {Processing ${\RR}_3$}.}
For each $rs\in {\RR}_3$,
compute the projection $\pi_\u(Z(rs))$ by Property $(1)$ of Lemma~\ref{Lemma:res3}.
Set ${\CC}_3=\{ \pi_\u(Z(rs))  \mid rs\in {\RR}_3\}$ 
and 
${\TT}_3=\{(C, \varnothing, 3)\mid C\in{\CC}_3\}$.

\noindent{\small \bf Comment.}
Since the zero sets of regular systems in ${\RR}$ are pairwise disjoint, 
after step $(3)$, $(4)$, $(5)$, we know that the element in ${\CC}_3$ has no
intersection with any element in ${\CC}_1$ or ${\CC}_2$. 
Note that it is possible that an element in ${\CC}_1$ has intersection 
with some element of ${\CC}_2$. 
So we need the following step to remove the common part between them.

\noindent{\small \bf Step $(6)$: {Merging}.}
Set ${\CC} =  {\CC}_1 \cup {\CC}_2 \cup {\CC}_3$
and
${\TT} =  {\TT}_1 \cup {\TT}_2 \cup {\TT}_3$.
Note that each element in ${\TT}$ is a triple 
$(C, {\PP}_C, {\II}_C)$, 
with $C\in{\CC}$ and 
where ${\II}_C$ is an integer of value $1, 2$ or $3$.
By means of the operation {\SMPD}, compute an intersection-free
basis ${\CC}'$ of ${\CC}$.
For each $C' \in {\CC}'$, compute ${\QQ}_{C'}$ (resp. ${\JJ}_{C'}$)
the union of the ${\PP}_C$ (resp. ${\II}_C$)
such that $C' \subseteq C$ holds.
Set  ${\TT}'=\{(C, {\QQ}_C, {\JJ}_C) \mid C\in {\CC}'\}$.

\noindent{\small \bf Step $(7)$: {Refinement}.}
To each $C\in{\CC}'$, apply operation {\MPD} to
the family of regular systems representing $C$, so as to obtain
another family ${\RR}_C$ of regular systems 
representing $C$ and whose zero sets are pairwise disjoint.
For each $rs\in{\RR}_C$, set ${\PP}_{rs}={\QQ}_C$ and
${\II}_{rs}={\JJ}_C$. 
Let ${\RR}'$ be
the union of the ${\RR}_C$, for all $C\in{\CC}'$.
Set ${\TT}''=\{(Z(rs), {\PP}_{rs}, {\II}_{rs}) \mid rs\in {\RR}'\}$.

\noindent{\small \bf Comment.}
 Recall that the union of zero sets of the $Z(rs)$, for all $rs\in{\RR}$
 equals $\KK^n$.
 Therefore, it follows from  Steps $(6)$ and $(7)$,
 that $\{Z(rs)\mid rs\in {\RR}'\}$ is a partition of $\KK^{n-1}$.

\noindent{\small \bf Step $(8)$: {Recursive call}.}
 Call ${\MakeCylindrical}({\RR}', n-1)$  to compute 
 a cylindrical decomposition ${\DD}'$ of $\KK^{n-1}$ 
 such that $Z(rs)$, for each $rs\in{\RR}'$, is
 a union of some cells of ${\DD}'$. 
 For each $D' \in {\DD}'$,
 observe that there exists 
 a unique $rs\in{\RR}'$ such that $D'\subseteq Z(rs)$,
 so set ${\PP}_{D'}={\PP}_{rs}$ and ${\II}_{D'}={\II}_{rs}$.
 Then, set
 ${\TT}'''=\{(D', {\PP}_{D'}, {\II}_{D'}) \mid D' \in {\DD}'\}$.

\noindent{\small \bf Comment.}
 By the comment below Step $(5)$, we know that for each triple 
 $(D', {\PP}_{D'}, {\II}_{D'})$ of ${\TT}'''$, 
 the values of ${\II}_{D'}$ 
 can only be 
 $\{1,2\}$, $\{2\}$ or
  $\{3\}$.
 Next, observe that for each $D' \in {\DD}'$ such that ${\II}_{D'}=\{2\}$
 or ${\II}_{D'}=\{3\}$ holds, we have ${\PP}_{D'}=\varnothing$,
 whereas for  each $D' \in {\DD}'$ such that ${\II}_{D'}=\{1,2\}$
 the set ${\PP}_{D'}$ is a nonempty finite family 
 of level $n$ polynomials in $\K[y_1,\ldots,y_n]$ 
 such that ${\PP}_{D'}$ separates above ${\DD}'$.
 In Step $(9)$ below,
 we lift the cylindrical decomposition ${\DD}'$ of $\KK^{n-1}$
 to a cylindrical decomposition ${\DD}$ of $\KK^{n}$. 

\noindent{\small \bf Step $(9)$: {Lifting}.}
 Initialize ${\DD}$ to the empty set.
 For each $D' \in {\DD}'$ such that ${\II}_{D'}=\{2\}$
 or ${\II}_{D'}=\{3\}$ holds, 
 let ${\DD}  :=  {\DD} \ \cup \ \{ D'\times\KK \}$.
 For each $D' \in {\DD}'$ such that ${\II}_{D'}=\{1,2\}$ holds,
 let ${\DD}={\DD}\cup\{D_p\}$, where
$$
D_p=\{(\alpha,y_n)\in\KK^n\mid \alpha\in D' \ \ {\rm and} \ \ ~p(\alpha,y_n)=0\},
$$
 for each $p\in{\PP}_{D'}$
 and let ${\DD}={\DD}\cup\{D_{*}\}$,
 where
$$
D_{*}= 
\{(\alpha,y_n)\in\KK^n\mid \alpha\in D'~\&~\left(\prod_{p\in{\PP}_{D'}}p(\alpha,y_n)\right)\neq0\},
$$
Finally, return ${\DD}$. 
The correctness of the algorithm follows from all the comments and
Definition~\ref{Defi:CylindricalDecomposition}.

\subsection{The Algorithm {\Partition} } %
\label{Sect:TheAlgorithmPartitionWholeSpace}

\noindent\underline{\small \bf Calling sequence.}  ${\Partition}(F,n)$

\noindent{\small \bf Input.} $F=\{f_1,\ldots,f_s\}$, a finite subset of $\K[y_1<\cdots<y_n]$.

\noindent{\small \bf Output.} A family {\RR} of regular systems, the zero sets
of which form an intersection-free basis of the
$s+1$ constructible sets $V(f_1),\ldots,V(f_s)$ 
and $\{y\in\KK^n \mid \left(\prod_{i=1}^s f_i(y)\right) \neq 0 \}$.

\noindent{\small \bf Step $(1)$:}  Let ${\BB}=$ {\SMPD}$(V(f_1),\ldots,V(f_s))$ be 
an intersection free basis of the $s$ constructible sets $V(f_1),\ldots,V(f_s)$.
For each element $B$ of ${\BB}$,
we apply operation {\MPD} to
the family of regular systems representing $B$ to compute
another family ${\RR}_B$ of squarefree regular systems 
such that the zero sets of regular systems in ${\RR}_B$ are 
pairwise disjoint and their union is $B$. Let ${\RR}$ be
the union of all ${\RR}_B$, $B\in{\BB}$. Clearly
the set $\{Z(rs)\mid rs\in {\RR}\}$ is an 
intersection-free basis of the $s$ constructible sets $V(f_1),\ldots,V(f_s)$.

\noindent{\small \bf Step $(2)$:} Let $f=\prod_{f_i\in F}f_i$ and $rs_{*}=[\varnothing,f]$. 
Set ${\RR}={\RR}\cup\{rs_{*}\}$. 
Obviously ${\RR}$ is the valid output.

\subsection{The Algorithm {\Decompose} } %
\label{Sect:TheAlgorithmCylindricalDecompose}

\noindent\underline{\small \bf Calling sequence.}  ${\Decompose}(F,n)$

\noindent{\small \bf Input.} $F$, a finite subset of $\K[y_1<\cdots<y_n]$.

\noindent{\small \bf Output.}  an $F$-invariant cylindrical 
                               decomposition of $\KK^n$.

\noindent{\small \bf Step $(1)$:} If $n>1$, go to step $(2)$. 
Otherwise let $\{p_1,\ldots,p_r\}$, $r\geq0$, be the
set of irreducible divisors of non-constant elements of $F$.
If $r=0$, set ${\DD}=\KK$ and exit.
Otherwise set 
$$
D_i=\{y_1\in\KK\mid p_i(y_1)=0\}, 1\leq i\leq r,
$$
and $D_{r+1}=\{y_1\in\KK\mid p_1(y_1)\cdots p_r(y_1) \neq 0\}.$
Clearly ${\DD}=\{D_i\mid 1\leq i\leq r+1\}$ is an $F$-invariant cylindrical decomposition of $\KK$.

\noindent{\small \bf Step $(2)$:} 
Let ${\RR}$ be the output of {\Partition}$(F,n)$.

\noindent{\small \bf Step $(3)$:}  
Call algorithm {\MakeCylindrical}$({\RR},n)$, to compute
a cylindrical decomposition ${\DD}$ of $\KK^n$ such that the zero set
of each regular system in ${\RR}$ is a union of some cells in ${\DD}$.
Clearly, ${\DD}$ is an intersection-free basis of the set $\{Z(rs)\mid rs\in {\RR}\}$,
which implies ${\DD}$ is an intersection-free basis of the
$s+1$ constructible sets $V(f_1),\ldots,V(f_s)$ and $\{y\in\KK^n \mid \left(\prod_{i=1}^s f_i(y)\right) \neq 0 \}$.
Therefore, ${\DD}$ is an $F$-invariant cylindrical decomposition of $\KK^n$.
\section{Cylindrical algebraic decomposition}
\label{sec:cad}
In this section, we show how to compute a
{\cad} of $\R^n$
from a cylindrical decomposition of $\C^n$.
This section starts with reviewing basic notions for {\cad}~\cite{Arnon84}.
A theorem (Theorem~\ref{Theorem:collins}) due to Collins~\cite{col75} is then reviewed, 
where the relation between complex and real roots of a polynomial with
real coefficients is shown. 
The bridge from cylindrical decomposition to {\cad} is built in Corollary~\ref{Corollary:stack},
which can be directly obtained from Collins' theorem.
The main algorithm ${\CAD}$ and its subroutines
are stated in four subsections.

A {\em semi-algebraic set}~\cite{bpr06} of $\R^n$ is a 
subset of $\R^n$ which can be written as a 
finite union of sets of
the form:
$$
 \{y\in\R^n\mid \forall f\in F, f(y)=0~\mbox{and}~\forall g\in G, g(y)>0\},
$$
where both $F$ and $G$ are finite subsets of the polynomial ring $\R[y_1,\ldots,y_n]$.

Given an $n$-dimensional real space $\R^n$, a nonempty connected subset
of $\R^n$ is called a {\em region}. 
For any subset $S$ of $\R^n$, a {\em decomposition} of $S$ is a finite collection
of disjoint regions whose union is $S$.
For a region $R$, the {\em cylinder}
over $R$, written $Z(R)$, is $R\times\R^1$.
Let $f_1<\cdots<f_r,r\geq0$ be continuous, real-valued functions 
defined on $R$. Let $f_0=-\infty$ and $f_{r+1}=+\infty$.
For any $f_i$, $1\leq i\leq r$, we call
the set of points $\{(a, f_i(a))\mid a\in R\}$ the {\em $f_i$-section} of $Z(R)$. 
For any two functions $f_i, f_{i+1}$, $0\leq i\leq r$, the set of points 
$(a, b)$, where $a$ ranges over $R$ and 
$f_i(a)<b<f_{i+1}(a)$, is called the {\em $(f_i, f_{i+1})$-sector} of $Z(R)$.
All the sections and sectors of $Z(R)$
can be ordered as 
$$(f_0,f_1)<f_1<\cdots<f_r<(f_r, f_{r+1}).$$
Clearly they form a decomposition of $Z(R)$, 
which is called a {\em stack} over $R$.

A decomposition ${\EE}$ of $\R^n$ is {\em cylindrical} 
if either $(1)$ $n=1$ and ${\EE}$
is a stack over $\R^0$, or $(2)$ $n>1$, and there is a cylindrical decomposition ${\EE}'$
of $\R^{n-1}$ such that for each region $R$ in ${\EE}'$, some subset of ${\EE}$ is a stack
over $R$. Moreover, We say that ${\EE}$ induces ${\EE}'$. 
A decomposition is {\em algebraic} if each of its regions is a semi-algebraic set.
A {\em cylindrical algebraic decomposition} of $\R^n$ is a decomposition which
is both cylindrical and algebraic.

Let $p$ be a polynomial of $\R[y_1,\ldots,y_n]$, and let $S$ be a subset of $\R^n$.
The polynomial $p$ is {\em invariant} on $S$ (and $S$ is $p$-invariant), 
if the sign of $p(\alpha)$ does not change when $\alpha$ ranges over $S$. 
Let $F\subset\R[y_1,\ldots,y_n]$ be
a finite polynomial set. We say $S$ is $F$-invariant if each $p\in F$ is
invariant on $S$. A cylindrical algebraic decomposition ${\EE}$ is $F$-invariant if 
$F$ is invariant on each region of ${\EE}$.

Let $p$ be a polynomial of $\R[y_1,\ldots,y_n]$, 
and let $R$ be a region in $\R^{n-1}$. $p$ 
is {\em delineable} on $R$ if the real zeros of $p$
define continuous real-valued functions 
$\theta_1,\ldots,\theta_s$ such that, for all $\alpha\in R$, 
$\theta_i(\alpha)<\cdots<\theta_s(\alpha)$,
and for each $\theta_i$ there is
an integer $m_i$ such that $m_i$ is the multiplicity of the root 
$\theta_i(\alpha)$ of $p(\alpha, y_n)$.
Note that if $k=0$, $V(p)$
has no intersection with $Z(R)$. 
Clearly when $p$ is delineable on $R$, its real zeros
naturally determine a stack over $R$. 

Let ${\EE}$ be a {\cad} of $\R^n$.
As suggested in \cite{Arnon84}, each region $e\in{\EE}$ can be
represented by a pair $(I,S)$, where $I$ is the {\em index} of $e$ and
$S$ is a {\em sample point} for $e$.
The index $I$ and the sample point $S$ of $e$ are defined as follows.
If $n=1$, let 
$$e_1<e_2<\cdots< e_{2m}<e_{2m+1}, m\geq0$$ 
be the elements of ${\EE}$. 
For each $e_i$, the index of $e_i$ is defined as $(i)$.
For each $e_i$, its sample point is any algebraic point belonging to $e_i$.
Let ${\EE}'$ be the {\cad} of $\R^{n-1}$ induced by ${\EE}$.
Suppose that region indices and sample points have
been defined for ${\EE}'$.
Let 
$$e_{i,1}<e_{i,2}<\cdots< e_{i,2m_i}<e_{i,2m_i+1}, m_i\geq0$$ 
be the elements of ${\EE}$ which form a stack
over the region $e_i$ of ${\EE}'$.
Let $(i_1,\ldots,i_{n-1})$ be the index of $e_i$.
Then the index of $e_{i,j}$ is defined as $(i_1,\ldots,i_{n-1}, j)$.
Let $S'$ be a sample point of $e_i$.
Then the sample point of $e_{i,j}$ is
an algebraic point belonging to $e_{i,j}$ such that its first $n-1$ coordinates
are the same as that of $S'$.

\begin{Theorem}[Collins]
\label{Theorem:collins}
Let $p$ be a polynomial of ring $\R[y_1<\cdots<y_n]$ and $R$ be a 
region of $\R^{n-1}$. If $init(p)\neq0$ on $R$ and 
the number of distinct complex roots of $p$ is invariant on $R$, then
$p$ is delineable on $R$.
\end{Theorem}
\begin{Corollary}
\label{Corollary:stack}
Let $F=\{p_1,\ldots,p_r\}$ be a finite set of polynomials 
in $\R[y_1<\cdots<y_n]$ of level $n$. 
Let $R$ be a region of $\R^{n-1}$. 
Assume that for every $\alpha\in R$,
$(1)$ the initial of each $p_{i}$ does not vanish at $\alpha$;
$(2)$ all $p_{i}(\alpha,y_n)$, $1\leq i\leq r$, as polynomials of $\R[y_n]$, are squarefree and coprime.
Then each $p_i$ is delineable on $R$ and 
the sections of $Z(R)$ belonging to different $p_i$ and $p_j$ are disjoint.
\end{Corollary}
Let $R$ and $F$ be defined as in the above corollary. 
Then clearly the real roots of all $p\in F$ are continuous functions on $R$
and they together determine a stack over $R$.
The algorithm {\GenerateStack}, described in Section~\ref{Sect:TheAlgorithmGenerateStack},  
is a direct application of the above corollary.

\subsection{Real Root Isolation}
\label{Sect:TheAlgorithmIsolateZeros}
Let $\alpha=(\alpha_1,\ldots,\alpha_{n})$ be an algebraic point of $\R^{n}$.
Each ${\alpha_i}$ as an algebraic number is a zero of a nonconstant squarefree 
polynomial $t_i(y_i)$ of $\Q[y_i]$. 
Let $T$ be the set of all $t_i(y_i)$. 
Clearly $T$ is a zero dimensional squarefree regular chain of $\Q[\y]$.
On the other hand, if $T$ is a zero-dimensional regular chain of $\Q[\y]$, 
any real zero of $T$ is an algebraic point of $\R^{n}$.
Therefore any algebraic point $\alpha$ of $\R^n$
can be represented by a pair $(T,L)$, where $T$ is a zero-dimensional squarefree regular chain of 
$\Q[\y]$ such that $T(\alpha)=0$ and 
$L$ is an isolating cube containing $\alpha$ but not other zeros of $T$.
The pair $(T,L)$ is called a {\em regular chain representation} of $\alpha$, 
which will be used to  represent a sample point of {\cad}.

Next we provide the specification of an algorithm called {\IsolateZeros} for isolating
real zeros of univariate polynomials with real algebraic number coefficients.
It is a subroutine of the algorithm {\small \sf NREALZERO} proposed in~\cite{xz06} 
for isolating the real roots of a zero-dimensional regular chain.

\noindent\underline{\small \bf Calling sequence.} 
${\IsolateZeros}(\alpha^{(n-1)}, F, n)$

\noindent{\small \bf Input.} $\alpha^{(n-1)}$ is a point of $\R^{n-1}$, $n\geq1$, with a regular chain representation $(T',L')$.
If $n=1$, $T'=\varnothing$ and $L'=\varnothing$.
$F=\{p_1,\ldots,p_r\}$ is a list 
of non-constant polynomials of $\Q[y_1,\cdots,y_n]$ of level $n$ satisfying that
$(1)$ for $p_i\in F$, $T'\cup\{p_i\}$ is a squarefree regular chain of  $\Q[y_1,\ldots,y_n]$;
$(2)$ all $p_i(\alpha^{(n-1)},y_n)$, $1\leq i\leq r$, 
as polynomials of $\R[y_n]$, are squarefree and coprime.

\noindent{\small \bf Output.} A pair $(N,\nu)$. Let $p=\prod_{i=1}^r p_i$. 
$N=(N_1,\ldots, N_m)$ is a list of intervals with
rational endpoints with $N_1<\cdots<N_m$ such that each $N_j$
contains exactly one real zero of $p(\alpha^{(n-1)},y_n)$. 
$\nu=(\nu_1,\ldots,\nu_m)$ is list of integers, where $1\leq \nu_i \leq r$,
such that the zero of $p(\alpha^{(n-1)}, y_n)$ in $N_j$ is a zero of $p_{\nu_j}(\alpha^{(n-1)}, y_n)$.

\subsection{The Algorithm {\GenerateStack}}
\label{Sect:TheAlgorithmGenerateStack}

\noindent\underline{\small \bf Calling sequence.} 
${\GenerateStack}(e', F, n)$

\noindent{\small \bf Input.} $e'$ is a region of a {\cad} ${\EE}'$ of $\R^{n-1}$, $n\geq1$, 
and $e'$ is represented by its index $I'$ and its sample point $S'$. 
Let $(T',L')$ be the regular chain representation of $S'$.
If $n=1$, $I',T',L'=\varnothing$.
$F$ is a finite set of polynomials 
in $\Q[y_1,\ldots,y_n]$ of level $n$. 
The region $e'$ and the polynomial set $F$ satisfy the conditions specified in 
Corollary~\ref{Corollary:stack}.

\noindent{\small \bf Output.} A stack ${\SS}$ over $e'$.

\noindent{\small \bf Step $(1)$.} 
If $F=\varnothing$, go to step $(2)$. 
Otherwise call algorithm ${\IsolateZeros}(S', F, n)$ to isolate the real roots of 
polynomials in $F$ w.r.t $y_n$ at the sample point $S'$ of $e'$. 
Let $(N,\nu)$ be the output.
If $N\neq\varnothing$, go to step $(3)$.

\noindent{\small \bf Step $(2)$.}
Let $I=(I',1)$. 
Let $T=T'\cup\{y_n\}$, $L=L'\times [0, 0]$, 
$S=(T, L)$ and return ${\SS}=((I,S))$.

\noindent{\small \bf Step $(3)$.}
Let $N_1=[a_1,b_1],\ldots,N_m=[a_m,b_m],$ $m>0$
be the elements of $N$.
For $1\leq i\leq 2m+1$, set $I_i=(I', i)$.
Let $s_1$ be the greatest integer less than $a_1$.
Let $s_{2m+1}$ be the smallest integer greater than $b_m$.
For $1\leq i\leq m-1$, let $s_{2i+1}=\frac{b_i+a_{i+1}}{2}$. 
For $0\leq i\leq m$, 
Let $T_{2i+1}=T'\cup\{y_n-s_{2i+1}\}$, $L_{2i+1}=L'\times [s_{2i+1}, s_{2i+1}]$ and set 
$S_{2i+1}=(T_{2i+1}, L_{2i+1})$.
For $1\leq i\leq m$, let $T_{2i}=T'\cup p_{\nu_i}$, $L_{2i}=L'\times N_i$ and
set $S_{2i}=(T_{2i}, L_{2i})$.
Finally, set ${\SS}$ be the list of all $(I_i, S_i)$, $1\leq i\leq 2m+1$. 
Then ${\SS}$ is the stack over $e'$.

\subsection{The Algorithm {\MakeSemiAlgebraic} } %
\label{Sect:TheAlgorithmMakeSemiAlgebraic}

\noindent\underline{\small \bf Calling sequence.} 
${ \MakeSemiAlgebraic({\DD}, n) }$

\noindent{\small \bf Input.} ${\DD}$ is a cylindrical decomposition of $\C^n$, $n\geq1$. 

\noindent{\small \bf Output.} A {\cad} ${\EE}$ of $\R^n$
such that, for each element $D$ of ${\DD}$, 
the set $D\cap\R^n$ is a union of some regions in ${\EE}$. 

\noindent{\small \bf Step $(1)$.} 
If $n>1$ go to $(2)$. 
Otherwise let $D_1,\ldots,D_r, D_{r+1}$, $r\geq0$ be the elements of ${\DD}$.
For each $1\leq i\leq r$, let $p_i$ be the polynomial such that
$D_i=\{y_1\mid p_i(y_1)=0\}$. 
Let ${\EE}$ be the output of $\GenerateStack(\varnothing,\{p_1,\ldots,p_r\}, 1)$.
Clearly ${\EE}$ is a {\cad} of $\R^1$.

\noindent{\small \bf Step $(2)$.}
Let ${\DD}'$ be the cylindrical decomposition 
of $\C^{n-1}$ induced by ${\DD}$. 
Call {\MakeSemiAlgebraic} recursively to compute a {\cad} ${\EE}'$ of $\R^{n-1}$.

\noindent{\small \bf Step $(3)$.}
In this step we lift the {\cad} ${\EE}'$ of $\R^{n-1}$
to ${\EE}$. Initialize ${\EE}=(~)$. 
For each region $e'$ of ${\EE}'$,
let $D'$ be the cell of ${\DD}'$ such that $e'\subset D'\cap\R^n$.
Let $D_1,\ldots, D_{r}, D_{r+1}$, $r\geq0$ be the cells of ${\DD}$ such that
$D'\times\C=\cup_{j=1}^{r+1}D_{j}.$
For each $1\leq j\leq r$, let $p_j$ be the polynomial such that
$D_j=\{(\alpha, y_n)\mid \alpha\in D' ~\&~ p_j(\alpha, y_n)=0\}.$
Add output of $\GenerateStack(e', \{p_1,\ldots,p_{r}\}, n)$ into ${\EE}$.
Clearly ${\EE}$ is a {\cad} of $\R^n$ and
for each $D\in{\DD}$, the set $D\cap\R^n$ is a union of some regions in ${\EE}$.

\subsection{The Algorithm {\sf CAD} } %
\label{Sect:TheAlgorithmCAD}

\noindent\underline{\small \bf Calling sequence.} 
{\sf \small CAD}$(F, n)$

\noindent{\small \bf Input.} $F$ is a finite subset of $\Q[y_1<\cdots<y_n]$, $n\geq1$. 

\noindent{\small \bf Output.} An $F$-invariant {\cad} ${\EE}$ of
$\R^n$.

\noindent{\small \bf Step $(1)$.} 
Let ${\DD}={\Decompose}(F,n)$ be an $F$-invariant
cylindrical decomposition of $\C^n$. 

\noindent{\small \bf Step $(2)$.} 
Call algorithm ${\MakeSemiAlgebraic}$
to compute a {\cad} ${\EE}$ of $\R^n$ such that, 
for each element $D$ of ${\DD}$, 
the set $D\cap\R^n$ is a union of some regions in ${\EE}$.
Since ${\DD}$ is an intersection-free basis of the
$s+1$ constructible sets $V_{\C}(f_1),\ldots,V_{\C}(f_s)$ and 
$\{y\in\C^n \mid \left(\prod_{i=1}^s f_i(y)\right) \neq 0 \}$, 
${\EE}$ is an intersection-free basis of the $s+1$
semi-algebraic sets  $V_{\R}(f_1),\ldots,V_{\R}(f_s)$ and 
$\{y\in\R^n \mid \left(\prod_{i=1}^s f_i(y)\right) \neq 0 \}$.
Note that each element in ${\EE}$ is connected. 
Therefore ${\EE}$ is an $F$-invariant cylindrical
algebraic decomposition of $\R^n$.

\section{Examples and Experimentation}
\label{sec:experimental}
\subsection{An Example}   %
\label{Sect:CDexample}
Let us illustrate our method by a simple and classical example.
Consider the parametric parabola $p = ax^2+bx+c$.
Set the order of variables as $x>c>b>a$.
The first step {\Partition} generates four regular systems,
 whose zero sets form a partition of $\C^4$.
$$
r_1 := \left\{
\begin{array}{rcl}
c&=&0\\
b&=&0\\
a&=&0
\end{array}
\right.,\ \
r_2 := \left\{
\begin{array}{rcl}
bx+c&=&0\\
b&\neq&0\\
a&=&0
\end{array}
\right.,
$$
$$
r_3 := \left\{
\begin{array}{rcl}
ax^2+bx+c&=&0\\
a&\neq&0
\end{array}
\right.,\ \
r_4 := \left\{
\begin{array}{l}
ax^2+bx+c\neq0
\end{array}.
\right.
$$
Next we trace the algorithm {\MakeCylindrical}.
Initialize the sets $\RR_1:=\{r_2, r_3\}$, $\RR_2:=\{r_4\}$ 
and $\RR_3:=\{r_1\}$.
Since $x$ appears in the equations
of $r_2$ and $r_3$, \SeparateZeros$(\RR_1)$ is called
to obtain a family of pairs
$$\{(C_1, \{t\}), (C_2, \{p\}), (C_3, \{q\})\},$$
defined as follows, 
which separates $Z(r_2)\cup Z(r_3)$.
$$
\begin{array}{lcl}
C_1:\{a=0, b\neq0\}&\rightarrow&\{t\}:\{bx+c\}\\
C_2:\{a(4ac-b^2)\neq0\}&\rightarrow&\{p\}:\{ax^2+bx+c\}\\
C_3:\{4ac-b^2=0, a\neq0\}&\rightarrow&\{q\}:\{2ax+b\}\\
\end{array}
$$
The projection of $Z(r_4)$
is the 
values such that $a$, $b$, $c$ do not
vanish simultaneously, denoted by $C_4$.
The projection of $Z(r_1)$
is the set $\{a=b=c=0\}$, denoted by $C_5$.

Note that $C_1,C_2,C_3$ are all subsets of $C_4$.
In the {\bf Merging} step, by calling {\SMPD}, 
we get another set $C_6:=\{a=b=0,c\neq0\}$ such
that $C_1,C_2,C_3, C_5$ and $C_6$ are pairwise disjoint
and their union is $\C^3$.
Moreover, for each $C_i$, there is a family of
polynomials and indices associated to it.
\begin{center}
\fbox{
\mbox{
\begin{tabular}{ccccc}
$C_1$    & $C_2$ & $C_3$   & $C_5$ & $C_6$\\
$\{t\}$ & $\{p\}$   & $\{q\} $  & $\varnothing$ & $\varnothing$\\
$\{1,2\}$ & $\{1,2\}$ & $\{1,2\}$ & $\{3\}$ & $\{2\}$
\end{tabular}}
}
\end{center}
Since each $C_i$ is already the zero set of some regular
system,
$$
{\MakeCylindrical}(\{C_1,C_2,C_3,C_5,C_6\},3)
$$
is called recursively to compute a cylindrical decomposition of $\C^3$.
By the {\bf Lifting} step, we finally obtain a
$p$-invariant cylindrical decomposition of $\C^4$.
Let $r=4ac-b^2$, the decomposition can be described by the following tree.

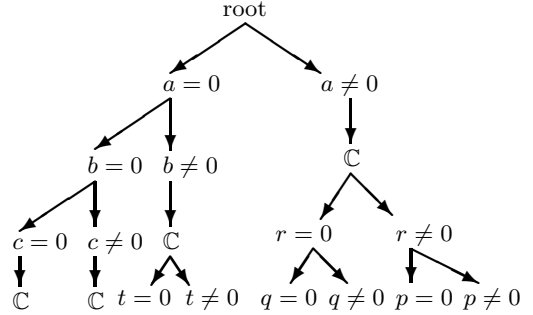
\setlength{\unitlength}{1cm}
\begin{picture}(8,4)\thicklines
\put(3.7,4.1){root}
\put(4,4){\vector(-3,-2){1}}
\put(4,4){\vector(3,-2){1}}
\put(2.9,3.1){$a=0$}
\put(5,3.1){$a\neq0$}
\put(3,3){\vector(-3,-2){1}}
\put(3,3){\vector(0,-1){0.7}}
\put(1.9,2){$b=0$}
\put(2.9,2){$b\neq0$}
\put(2,1.9){\vector(-3,-2){1}}
\put(2,1.9){\vector(0,-1){0.6}}
\put(3,1.9){\vector(0,-1){0.6}}
\put(0.9,1){$c=0$}
\put(1.9,1){$c\neq0$}
\put(2.9,1){$\C$}
\put(1,0.9){\vector(0,-1){0.4}}
\put(2,0.9){\vector(0,-1){0.4}}
\put(3,0.9){\vector(-2,-3){0.25}}
\put(3,0.9){\vector(2,-3){0.25}}
\put(0.9,0.2){$\C$}
\put(1.9,0.2){$\C$}
\put(2.3,0.25){$t=0$}
\put(3.2,0.25){$t\neq0$}
\put(5.4,3){\vector(0,-1){0.6}}
\put(5.3,2.1){$\C$}
\put(5.4,2){\vector(-2,-3){0.4}}
\put(5.4,2){\vector(1,-1){0.55}}
\put(4.4,1.1){$r=0$}
\put(6.0,1.1){$r\neq0$}
\put(4.9,1){\vector(-2,-3){0.3}}
\put(4.9,1){\vector(1,-1){0.45}}
\put(4.2,0.25){$q=0$}
\put(5.1,0.25){$q\neq0$}
\put(6.2,1){\vector(0,-1){0.45}}
\put(6.2,1){\vector(2,-1){0.9}}
\put(6.0,0.25){$p=0$}
\put(6.9,0.25){$p\neq0$}
\end{picture}

To compute a $p$-invariant {\cad} of $\R^4$
from the above tree is straightforward.
Starting from the root, one first obtains 
the trivial zero $0$ of $a$, 
which decomposes $a\neq0$ into two connected
cells $a<0$ and $a>0$. The real line
is thus divided into three parts. 
For each part, one then substitutes its sample point
into its children which are equations, from where
one can determine the sample points for
the children which are inequations.
Continuing in this manner,
one finally obtains
a {\cad} of $\R^4$ with $27$ cells.
The number of cells is the same as that obtained in~\cite{brown01}. 
In fact, it is the minimal number of cells
one can obtain for a $p$-invariant {\cad} of $\R^4$.
\subsection{Experimental Results}
In this section, we present experimental results
obtained with an implementation of 
the algorithms presented in this paper.
Our code is in {\Maple} 12 running on a  
computer with Intel Core 2 Quad {\small CPU} (2.40{\small GHz}) and 3.0{\small GB} total memory.
The test examples, listed in appendix for the reader's convenience, 
 are taken from diverse 
papers~\cite{Dolzmann04,Arnon84,ch91,scott88,brown01,Collins02,collins98} on {\cad}.
The time-out for a test run is set to 2 hours.

In Table 1, we show the total computation time of
{\CAD} and the
time spent on three main phases of it,
which are
{\Partition} ({\small \sf Partition} for short), 
{\MakeCylindrical} ({\small \sf M.C.} for short) and 
{\MakeSemiAlgebraic} ({\small \sf M.S.A.} for short).
We also report the number of elements ($N_\R$) in the {\cad}.
Aborted computations due to time-out are marked with ``-''.
From the table, one can see that, 
except examples $14$ and $16$, the steps of the algorithm
dedicated to computations in complex space
dominate the step taking place in the real space.

\begin{figure}[h]
\centering
{\small
\begin{tabular}{cccccc} 
Sys  & {\small \sf Partition} & {\small \sf M.C.}  & {\small \sf M.S.A.} & Total & $N_\R$ \\\hline
   1 &  0.024   & 0.096    & 0.024  & 0.144  & 27\\
   2 &  1.184   & 2.856    & 1.048  & 5.088 &  895\\
   3 &  0.004   & 7.512    & 0.704  & 8.220 & 233\\
   4 &  0.264   & 1.368    & 1.080  & 2.716 & 421\\
   5 &  0.016   & 0.052    & 0.116  & 0.184 & 55\\
   6 &  0.108   & 0.156    & 0.120  & 0.384 & 41\\
   7 &  2.704   & 3.600    & 1.360  & 7.664 & 893\\
   8 &  0.380   & 1.608    & 1.196  & 3.184 & 365\\
   9 &  0.288   & 0.532    & 0.264  & 1.084 & 209\\
  10 &  5.668   & 48.079   & 18.833 & 72.640 & 3677\\
  11 &  0.252   & 1.192    & 0.620  & 2.068 & 563\\
  12 &  2.664   & 135.028  & 88.142 & 225.862 & 20143\\
  13 &  10.576  & 35.846   & 6.905  & 53.335 & 4949\\
  14 &  5.728   & 71.760   & 2520.354 & 2597.878  & 27547\\
  15 &  690.731 & 2513.817 & 299.250 & 3503.954 & 66675\\
  16 &  895.435 & 2064.469 & - & - &- \\
  17 &  0.052 & - & - & - &- \\
  18 &  - & - & - & - &- \\\hline
\end{tabular}}
\medskip
\newline
{\textbf{Table 1} Timing (s) and number of cells for {\CAD}}
\end{figure}

In Table 2, we show the total computation time 
of the algorithm {\Decompose} ({\small \sf C.D.} for short)
and the time spent on three main operations of it, which are
respectively  {\SeparateZeros}({\small \sf Separate} for short), {\MPD}
and {\SMPD}. We can see that the cost of algorithm {\Decompose}
is dominated by {\SMPD}. The number of
elements ($N_\C$) in the cylindrical decomposition of ${\C}^n$ 
is also reported.

\begin{figure}[h]
\centering
{\small
\begin{tabular}{cccccc} 
Sys  & {\small \sf Separate} & {\MPD}  & {\SMPD} & Total & $N_\C$\\\hline
   1 & 0.020         & 0.012         & 0.084         &  0.156 & 8\\
   2 & 0.508         & 0.252         & 2.268         &  4.052 & 63\\
   3 & 3.856         & 0.836         & 2.460         &  7.880 & 24\\
   4 & 0.280         & 0.088         & 1.036         &  1.648 & 65\\
   5 & 0.032         & 0.008         & 0.012         &  0.064 & 7\\
   6 & 0.036         & 0.012         & 0.092         &  0.268 & 13\\
   7 & 1.100         & 0.652         & 2.416         &  6.320 & 58\\
   8 & 0.536         & 0.144         & 1.040         &  2.008 & 55\\
   9 & 0.120         & 0.032         & 0.384         &  0.816 & 26\\
   10 & 3.204        & 0.756         & 49.031        &  54.119 & 594\\
   11 & 0.128        & 0.032         & 0.960         &  1.416 & 49\\
   12 & 8.508        & 2.024         & 125.104       &  138.188 & 856\\
   13 & 2.040        & 1.784         & 42.578        &  47.002 & 407\\
   14 & 5.741        & 2.092         & 64.875        &  76.956 & 983\\
   15 & 83.469       & 62.736        & 3066.071      &  3232.073 &2974\\
   16 & 66.516       & 377.664       & 2501.947      &  2959.904   & 5877\\
\hline
\end{tabular}}
\medskip
\newline
{\textbf{Table 2} Timing (s) and number of cells for {\small \sf C.D.}}
\end{figure}

The data reported in two tables shows that {\SMPD} is the
dominant operation, which computes intensively  {\small GCD}s
of polynomials modulo regular chains.
This suggests that the modular methods
and efficient implementation techniques in~\cite{DMSWX05a,LMS07,LPM09}
(use of FFT-based polynomial arithmetic, \ldots)
have a large potential 
for improving the implementation of our {\cad} algorithm.

\section{Conclusion}
We have presented a new approach for computing 
cylindrical algebraic decompositions.
Our main motivation is to understand
the relations between {\cad}s and triangular decompositions,
studying how the efficient techniques developed
for the latter ones can benefit to the former ones.
 
Our method can be applied for solving {\QE} problems directly.
However, to solve practical problems efficiently, 
our method needs to be equipped with existing
techniques, like partially built {\cad}s, 
for utilizing the specific feature of input
problems. Such issues will be addressed in a future paper.

\appendix

1. Parametric parabola\\
$\{ax^2+bx+c\},x>c>b>a.$

2. Whitney umbrella\\
$\{x-uv, y-v, z-u^2\},v>u>z>y>x.$

3. Quartic\\
$\{x^4+px^2+qx+r\},x> p> q> r.$

4. Sphere and catastrophe\\
$\{z^2+y^2+x^2-1, z^3+xz+y\},x> y> z.$

5. Arnon-84\\
$\{y^4-2y^3+y^2-3x^2y+2x^4\},y> x.$

6. Arnon-84-2\\
$\{144y^2+96x^2y+9x^4+105x^2+70x-98,$\\ 
$xy^2+6xy+x^3+9x\},  y> x.$

7. A real implicitization problem\\
$\{x-uv, y-uv^2, z-u^2\},  v> u>z>y>x.$

8. Ball and circular cylinder\\
$\{x^2+y^2+z^2-1, x^2+(y+z-2)^2-1\}, z> y> x.$

9. Termination of term rewrite system\\
$\{x-r, y-r, x^2(1+2y)^2-y^2(1+2x^2)\}, r> x> y.$

10. Collins and Johnson\\
$\{3a^2r+3b^2-2ar-a^2-b^2,$\\ 
$3a^2r+3b^2r-4ar+r-2a^2-2b^2+2a,$\\ 
$a-1/2, b, r, r-1\}, r> a> b.$

11. Range of lower bounds\\
$\{a, az^2+bz+c, ax^2+bx+c-y\},$ \\
$z> c> b> a> x> y.$

12. $X$-axis ellipse problem\\
$\{b^2(x-c)^2+a^2y^2-a^2b^2,$ \\
 $x^2+y^2-1\}, y> x> b> c> a.$

13. Davenport and Heintz\\
$\{a-d, b-c, a-c, b-1, a^2-b\}, a> b> c> d.$

14. Hong-90\\
$\{r+s+t, rs+st+tr-a, rst-b\},$ \\
   $t> s> r> b> a.$

15. Solotareff-3\\
$\{r, r-1, u+1, u-v, v-1,$ \\
$3u^2+2ru-a, 3v^2+2rv-a,$ \\
$u^3+ru^2-au+a-r-1,$ \\
$v^3+rv^2-av-2b-a+r+1\},$ \\
$b> u> v> r> a.$

16.  Collision problem\\
$\{\frac{17}{16}t-6, \frac{17}{16}t-10,x-\frac{17}{16}t+1,$\\ 
$x-\frac{17}{16}t-1, y-\frac{17}{16}t+9,y-\frac{17}{16}t+7,$\\
$(x-t)^2+y^2-1\}, t> x> y.$

17. McCallum trivariate random  polynomial\\
$\{(y-1)z^4+xz^3+x(1-y)z^2+(y-x-1)z+y\},$\\
$z > y >x.$

18. Ellipse problem\\
$\{b^2(x-c)^2+a^2(y-d)^2-a^2b^2, a, b,  x^2+y^2-1\},$ \\
$y > x > d > c > b > a.$

\end{document}